\documentclass[twocolumn,showpacs,preprintnumbers,amsmath,amssymb,superscriptaddress, prl,floatfix]{revtex4-1}

\usepackage{graphicx}
\usepackage{dcolumn}
\usepackage{tabularx}
\usepackage{bm}
\usepackage{epstopdf}
\usepackage{amsmath}
\usepackage{lipsum}
\usepackage{color}
\usepackage{here}
\usepackage{float}

\begin{document}

\title{The Effect of a Noisy Driving Field on a Bistable Polariton System }

\author{H. Abbaspour}
\email[E-mail: ]{hadis.abbaspour@epfl.ch}

\author{G. Sallen}
\affiliation{Laboratory of Quantum Optoelectronics, \'Ecole Polytechnique F\' ed\'erale de Lausanne, CH-1015, Lausanne, Switzerland}
\author{S. Trebaol}
\affiliation{UMR FOTON, CNRS, Université de Rennes 1, Enssat, F22305 Lannion, France}
\author{F. Morier-Genoud}
\affiliation{Laboratory of Quantum Optoelectronics, \'Ecole Polytechnique F\' ed\'erale de Lausanne, CH-1015, Lausanne, Switzerland}
\author{M. T. Portella-Oberli}
\affiliation{Laboratory of Quantum Optoelectronics, \'Ecole Polytechnique F\' ed\'erale de Lausanne, CH-1015, Lausanne, Switzerland}
\author{B. Deveaud}
\affiliation{Laboratory of Quantum Optoelectronics, \'Ecole Polytechnique F\'ed\'erale de Lausanne, CH-1015, Lausanne, Switzerland}

\date{\today}

\begin{abstract}
We report on the effect of noise on the characteristics of the bistable polariton emission system. The present experiment provides a time resolved access to the polariton emission intensity. We evidence the noise-induced transitions between the two stable states of the bistable polaritons. It is shown that the external noise specifications, intensity and correlation time, can efficiently modify the polariton Kramers time and residence time. We find that there is a threshold noise strength that provokes the collapse of the hysteresis loop. The experimental results are reproduced by numerical simulations using Gross-Pitaeviskii equation driven by a stochastic excitation.  
\end{abstract}


\maketitle
\section{ I. INTRODUCTION}
Optical bistable systems share the characteristic to present two possible output steady states for the same input light intensity. This is observed as a hysteresis cycle in the output versus light intensity plot. Optical bistability was proposed by \cite{Szoeke1969, Lugiato1978, Abraham1982} and since then observed in many different systems such as cavity lasers \cite{Roy1980, Arimondo1983, Agrawal1984, Joshi2003}, atomic systems \cite{Joshi2003, Joshi2003a}, semiconductors \cite{Sfez1990, Asadpour2014, Gibbs1979}, and microcavity polaritons \cite{Baas2004, Paraiso2011}. Bistability is commonly used for device applications \cite{Abraham1982, Miller1984,Gibbs2012, Takase2014}, particularly in polariton systems as spin switch and optical memory \cite{Amo2010, Cerna2013}, optical transistor \cite{Ballarini2013}, laser \cite{Grosso2014}, and logic functions \cite{Liew2008}. The fidelity of the devices depends on the stability of the system. 

Nonlinearity and optical feedback constitute the basis of optical bistability. In a bistable regime, instabilities can play an important role in the output signal. In fact, nonlinearity couple with feedback fluctuations may induce transitions between the two stable states. Therefore, a noisier bistable system can give stochastic fluctuations in the output power. The characteristic time scale for these fluctuations is given by the transition probability defined by the residence time \cite{Gammaitoni1989, Johne2009} and Kramers time \cite{Kramers1940}, which depend on the noise strength and correlation time \cite{Johne2009}. Moreover, bistable systems in the presence of fluctuations may appear as a discriminator instead of a hysteresis loop. In order to get insight in the actual bistable behaviour of the system, which is related to the temporal variation of the population, we need to investigate the time behaviour of the output intensity. Likewise, this time resolved photon statistics might well reveal the genuine instability of the system, which should be considered in the realization of devices. Then, conversely, bistable optical systems can also provide an opportunity to investigate the fluctuations related to the nonlinear system.
\begin{figure*}[tb] 
\centering 
\includegraphics[width=\textwidth,height=5cm]{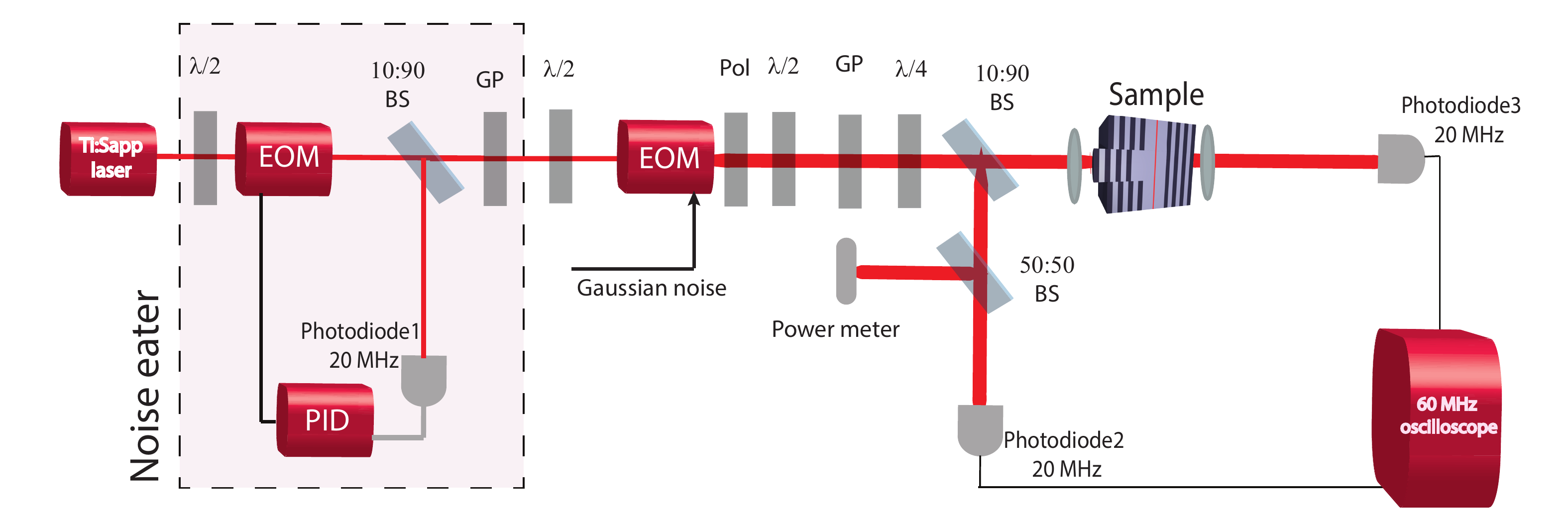} 
\caption[Experimental setup.] {\textbf{Experimental setup} Microcavity sample is excited using the cw Ti: Sapphire laser. Noise eater: the laser noise intensity is reduced to 0.02 percent through the first electro-optic modulator and the feedback loop (photodiode 1 and PID). Second EOM imprints 500 kHz Gaussian noise on the laser intensity. Excitation and transmission light are detected with two fast photodiodes. BS, Pol, GP, $\lambda/2$, $\lambda/4$ represent beam splitter, polarizer, Glan polarizer, half-wave plate and quarter-wave plate, respectively.}
\end{figure*}

Microcavity exciton polaritons are quasiparticles that arise from the strong coupling between excitons and photons \cite{Weisbuch1992}. They show a nonlinear behaviour due to polariton interactions coming from their excitonic content, and their properties can be accessed through the emitted photons. Indeed, polariton repulsive interactions produce a positive nonlinear feedback mechanism when using a detuned exciting beam, which can be observed as a hysteresis cycle in the output light intensity \cite{Baas2004, Paraiso2011}. The relative energy positions of the input laser and of the polariton state play an important role in the polariton energy shift, which leads to optical bistability regimes with different hysteresis widths. 

If the exciting laser is free of noise, at a given excitation power polaritons will stay either in the lower or in the upper branch of the hysteresis where they are stable. Otherwise, in the presence of noise and when the input laser power is within the bistable cycle, polaritons can escape from one of the stable branch to the other. The average time that the system stays in one bistable state before escaping to the other is defined as residence time. Then with a certain amount of noise strength polaritons undergo noise-assisted transitions between the steady states, and with a characteristic residence time they jump randomly up and down due to their coupling to the fluctuations. These fluctuations can originate from either external sources (the driving laser) or internal sources as a consequence of polariton with phonon and reservoir interactions. It has been theoretically predicted that the width of the measured hysteresis depends on the strength of the noise \cite{Johne2009}. Therefore, time-resolved analysis of the output emission is crucial to enlighten the effect of noise in bistable polariton system.

In this paper, we report the characteristics of polariton bistable systems by applying a controllable noise strength.  We demonstrate the effect of noise on the characteristics of a bistable polariton system. We show that the presence of noise induces a reduction of the width of the hysteresis cycle within the residence time scale. Moreover, the results reveal the noise threshold for which the hysteresis loop collapses and appears like a discriminator. The investigation is pursued by the study of the emitted photon statistics through time resolved measurements. We give evidence that the polariton emission intensity fluctuates between the lower and upper states of the polariton bistability due to noise-induced transitions. We determine the lower and upper state residence times as function of noise and excitation power. We determine the power for which the residence times of both branches are the same, which identifies the polariton Kramers time and we measure the Kramers times as a function of noise strength. Numerical simulations using Gross-Pitaeviskii equation driven by a stochastic excitation reproduce  very well the experimental results.

The paper is organized as follows: In Section II, we describe the sample and the experiments. Section III reports on the bistability measurements and time-resolved experimental studies. Section IV is dedicated to the theoretical model based on Gross-Pitaevskii equation driven by stochastic excitation. We give our conclusions in Section V.
\section{II. EXPERIMENTAL METHOD}
The sample is a single 8nm $In_{0.04}Ga_{0.96}As$ quantum well imbedded in a GaAs $\lambda$ microcavity with AlAs/GaAs distributed Bragg mirrors \cite{Kaitouni2006}. The Rabi splitting is 3.5 meV at zero detuning. Polaritons are confined in a patterned 3 $\mu$m diameter mesa engineered on top of the spacer layer. The experiments are performed with the sample held at a temperature of 4 K and at a cavity-exciton detuning of $\delta$=-1.15 meV. The linewidth of the zero-dimensional polariton ground state is $\gamma=70 \mu$eV. This allows to perform experiments with the single ground state level properly isolated from the other states. The experimental set-up is shown in Figure 1. We use a linearly polarized single mode cw Ti:Sapphire laser with 10 MHz linewidth and 2 $\%$ noise intensity standard deviation. Using the noise eater, we reduce this value to 0.02 $\%$. Then, the laser beam passes through an electro-optic modulator, which imprints on the DC laser power a controllable noise intensity with a 500 kHz bandwidth which is indeed much slower than the polariton gas frequency dynamics. We present the noise strengths as normalized standard deviation compared to the bistability width $\Delta$B measured without any additional noise on the laser. We vary the laser power, with a resolution equal to 1 $\mu$W, through a motorized rotating half-wave plate and a Glan polarizer. The sample is then excited at normal incidence with the laser circularly polarized to ensure polariton repulsive interactions and therefore polariton energy blueshift \cite{Takemura2014}. The laser has a spot diameter of 20 $\mu$m and is detuned of $\Delta$=0.4 meV above the ground state energy, consequently in the excitation bistability condition $\Delta>\sqrt{3}\gamma$ \cite{Baas2004}. The excitation and transmitted light are detected with 20 MHz bandwidth photodiodes. The time-resolved measurements are performed with a 60 MHz bandwidth oscilloscope.
\section{III. Experimental results}
\subsection{A.1. Effect of noise on polariton bistability}
\begin{figure}[tb] 
\centering 
\includegraphics[width=1\columnwidth]{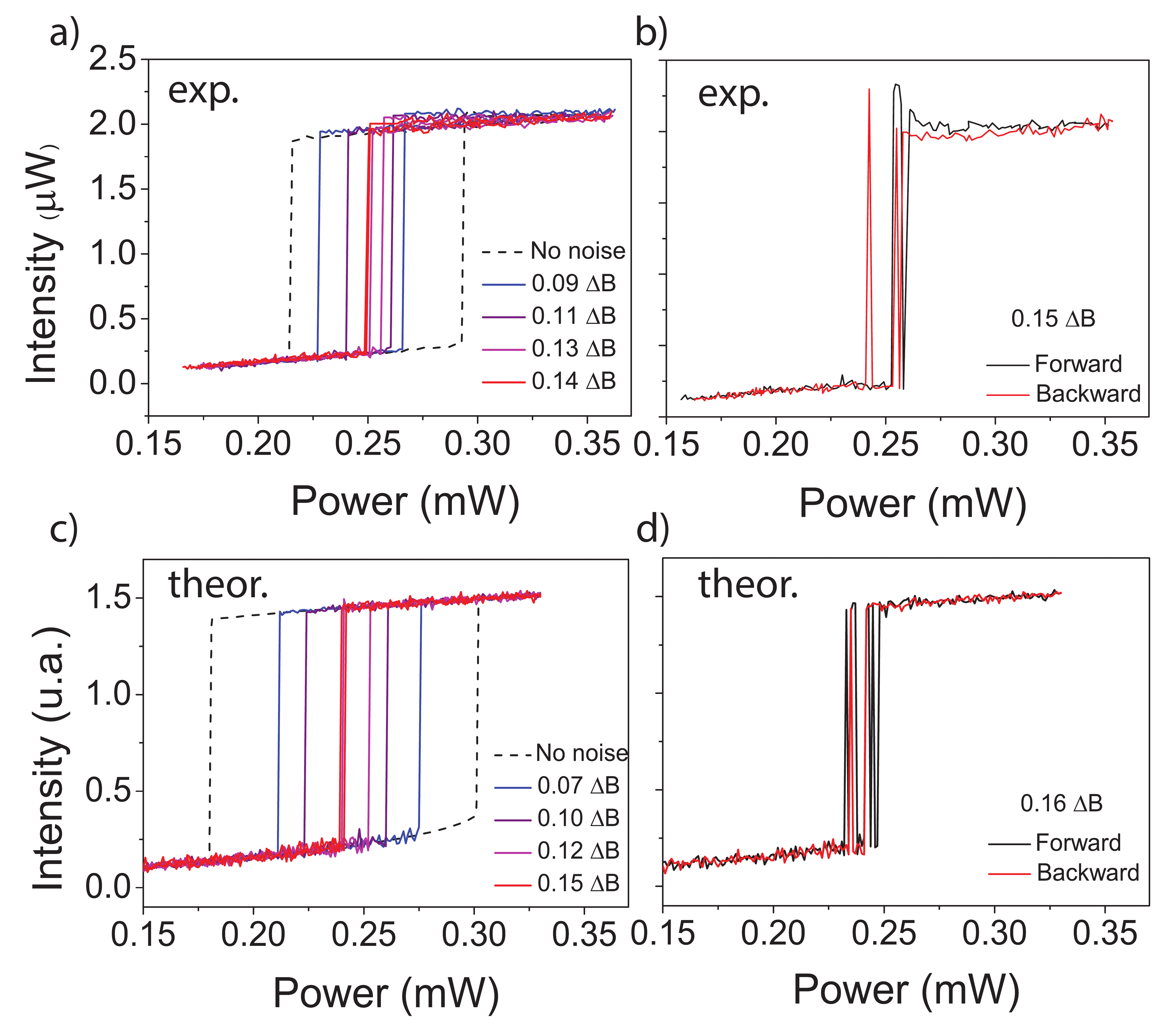} 
\caption[Zero-dimensional polariton bistability affected by the external Gaussian noise. ]{\textbf{Zero-dimensional polariton bistability affected by the external Gaussian noise.} \textbf{a} The initial experimental polariton bistability (dash line) and polariton bistabilities for different amounts of the laser noise: D=0.095 $\Delta$B, 0.11 $\Delta$B, 0.13 $\Delta$B and 0.14 $\Delta$B. Fluctuations on the applied laser intensity make the polariton bistability unstable. We measure the collapse of the optical hysteresis. \textbf{b} For larger amount of the laser noise (D=0.15 $\Delta$B) we start to observe transition between two states. \textbf{c} The initial theoretical polariton bistability (dash line) and polariton bistabilities for different amounts of the laser noise: D=0.07$\Delta$B, 0.10 $\Delta$B, 0.12 $\Delta$B and 0.15 $\Delta$B. \textbf{d} For a larger amount of the laser noise (D=0.16 $\Delta$B) we start to observe transitions between two states. The parameters used for the GPE are $\gamma_{p}$=0.08 meV, $\alpha_{1}$=0.34 meV and $\Delta$=0.4 meV. $\Delta$B is the width of the hysteresis without applying any noise.}
\end{figure}
\begin{figure}[b!] 
\centering 
\includegraphics[width=1\columnwidth]{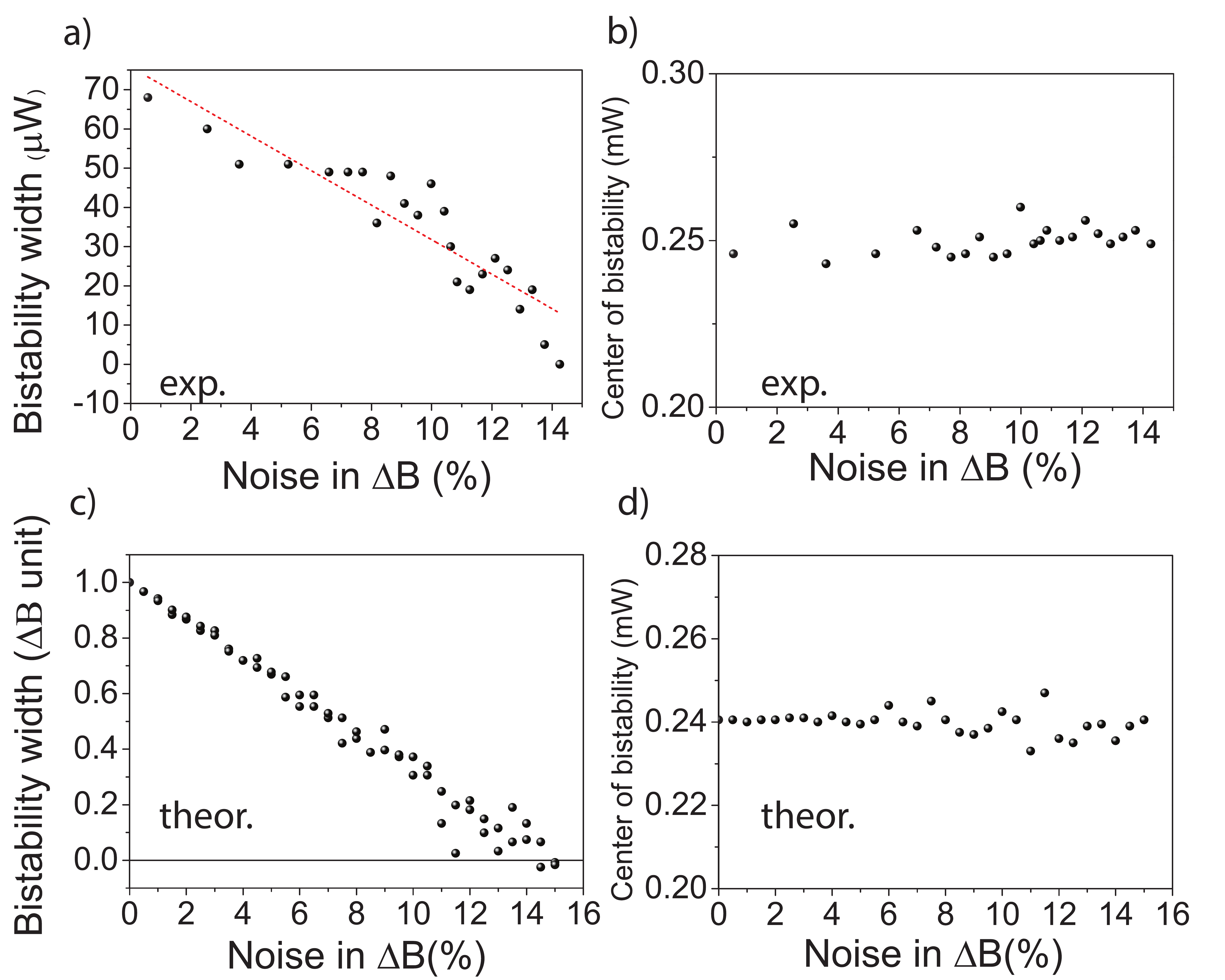} 
\caption[Characteristics of polariton bistability for different amounts of the external noise.]{\textbf{Characteristics of polariton bistability for different amounts of external noise.} \textbf{a} Experimental bistability width versus intensity of laser fluctuations. By increasing noise power bistability decreases and for D=0.14 $\Delta$B it is equal to zero.  \textbf{b} For different amplitude of laser fluctuations the middle point of each hysteresis fluctuates around 0.245 mW. \textbf{c} Theoretical bistability width (normalized to the maximum theoretical $\Delta$B) versus noise power in $\Delta$B unit. By increasing the noise power bistability decreases and for D=0.15 $\Delta$B it is equal to zero.  \textbf{d} For different amplitudes of the laser fluctuations, the middle point of each hysteresis loop fluctuates around 0.24 mW.}
\end{figure}
In what follows, we present both experimental and theoretical results, however all theoretical part will be explained in section IV. First we measure the polariton emission intensity as function of input laser power. The variation of the input power in the forward and backward directions builds the bistability curve in the output power. We repeat the measurements for a wide range of applied noise strengths. In Figure 2 (a), we display a set of bistability curves obtained for different amounts of applied noise. The bistability (black dashed) measured without any additional noise on the laser power has a hysteresis cycle width of $\Delta$B=68 $\pm$ 8 $\mu$W. The uncertainty on the bistability width originates from unavoidable experimental mechanical noise and also internal noise sources. We observe that the increase in the noise strength results in a decrease in the hysteresis width. Small amounts of noise perturb the emission behaviour close to the bistability thresholds, inducing a single transition to the other branch in which the system stays stable. This reduces the width of the hysteresis cycle.  Upon a certain amount of noise (here about 0.14 $\Delta$B) the hysteresis is screened and the system is seen as an optical discriminator at input power 0.245 $\pm$ 0.008 mW. Notice, for larger noise strengths, instability appears in the hysteresis cycle and random jumps occur between lower and upper branches (Fig. 2(b)). This evidences that although the system behaves always as a bistable system, but the intensity noise can bring instability for device applications.

In Figure 3 (a), we plot the measured hysteresis width as a function of the applied noise strength, and in Figure 3 (b) we display the input power corresponding to the middle position of the respective hysteresis loops. This result reveals a linear dependence of the bistability width with noise. In addition, the reduction of the hysteresis loop evolves symmetrically to collapse in the center position. By extrapolating the straight line we obtain the maximum width for the system free of any applied noise to the laser ($\Delta B \approx 76 \mu W$). It is worth mentioning that the real maximum bistability width can be even larger than this value. This difference originates from the minimal amount of noise which might be due to either external low frequency mechanical or thermal vibrations or even intrinsic noise in the polariton system.  
\begin{figure}[h!] 
\centering 
\includegraphics[width=1\columnwidth]{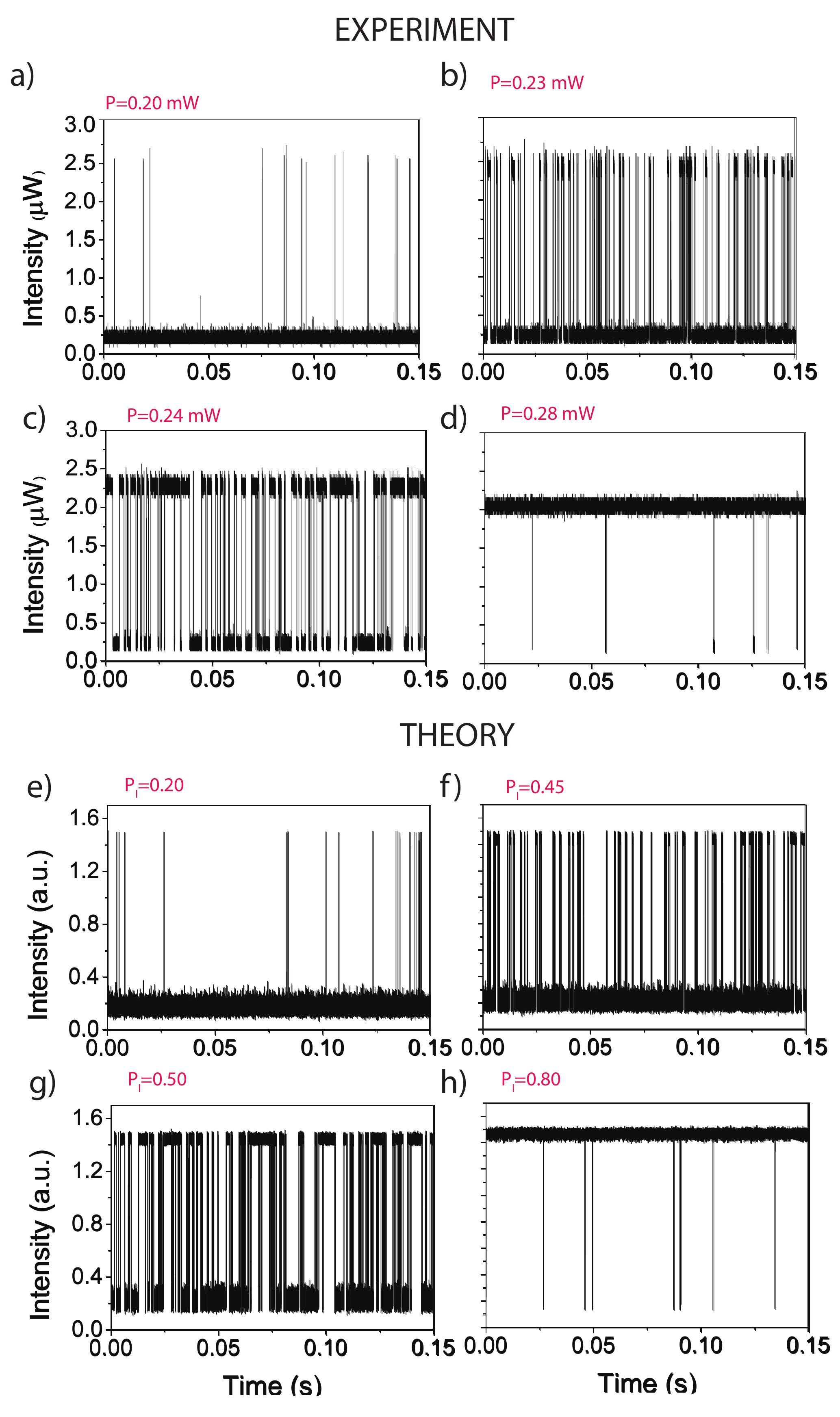} 
\caption[A single realization of polariton transitions between two stable states.]{\textbf{A single realization of polariton transitions between two stable states.} \textbf{a, b} For laser power below the middle of bistability (P=0.20, 0.23 mW) the lower state is the more stable state. \textbf{c} For the laser power around the middle of bistability (P=0.24 mW) polariton transitions between the two stable states occurs with the same probability. \textbf{d} For the laser power around the upper threshold (P=0.28 mW), the upper branch is the more stable state and we observe some random jumps to the lower state. The applied noise strength is of D=0.30 $\Delta$B. \textbf{e, f} For laser power below the middle of theoretical bistability ($P_{I}$=0.20, 0.45) the lower state is the more stable state. $P_{I}$ is defined in equation 4. \textbf{g} For a laser power close to the middle of the theoretical bistability loop ($P_{I}$=0.5), transitions between the two stable polariton states occur with the same probability. \textbf{h} For a laser power around the upper threshold ($P_{I}$=0.8), the upper branch is the most stable state and we observe some random jumps to the lower state.}
\end{figure}
\begin{figure}[h!] 
\centering 
\includegraphics[width=0.9\columnwidth]{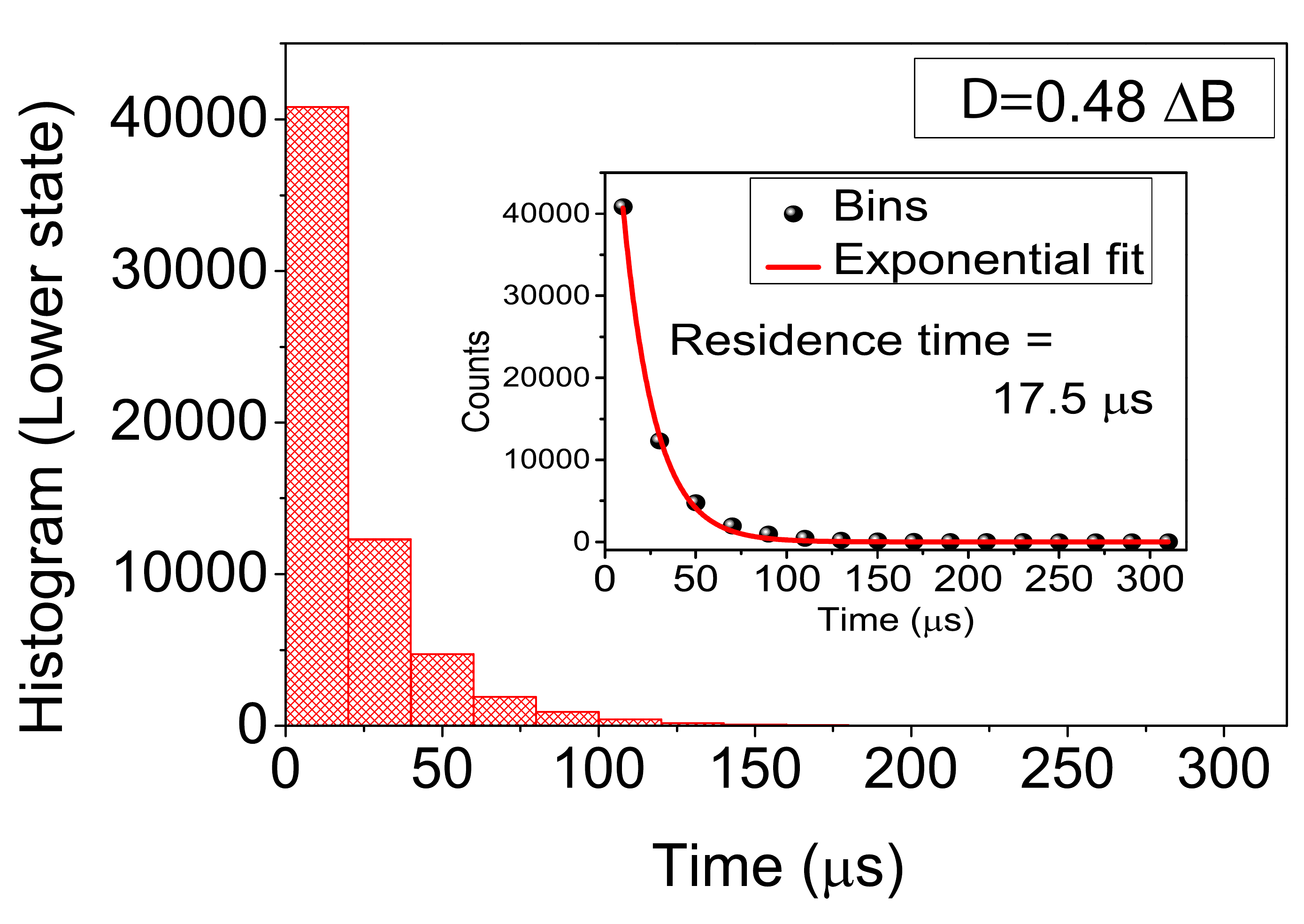} 
\caption[Residence time probability distribution.]{\textbf{Residence time probability distribution.} Histogram of the lower state residence time distribution for the noise intensity D=0.48 $\Delta$B at P=0.24 mW. Inset: Histogram of the distribution showing the exponential decay of 17.5 $\mu$s.}
\end{figure}

To understand the dependence of the hysteresis cycle with noise, it is very important to know how the bistable polariton system behaves. Note that in a bistable polariton system, the intensity of polariton emission is governed by both the polariton lifetime and polariton-polariton interactions, which strongly depend on the polariton population. In addition, the emission intensity depends on the spectral overlap between the laser field and the polariton state. Within the hysteresis cycle, in the forward direction, we observe that the emission intensity in the lower branch increases linearly with the laser power (Fig. 2 (a)). This is due to the fact that the laser and polariton energy overlap is very small and the emission intensity increases due to the enhancement of the polariton population with laser power. The increase of the polariton population activates the nonlinear polariton interaction, as a consequence the polariton energy blueshifts with laser power. However, this supply mechanism competes with the polariton lifetime loss mechanism. At a given threshold power, the feedback mechanism becomes dominant and polaritons reach a population at which interactions lock the polariton energy in resonance with the laser. This results in a better polariton coupling to the laser. This occurs at the upper power threshold of the hysteresis cycle, when the emission intensity jumps to the upper branch. In the backward direction, by decreasing the laser power, the emission remains in the upper branch until reaching the lower power threshold. This occurs at lower power because the polariton state is clamped to the laser and therefore both the polariton population and the nonlinearity strength are also clamped. At this lower threshold the losses due to polariton lifetime dominate the feedback mechanism, which reduces the polariton population. As a consequence, the nonlinear polariton interaction decreases and accordingly the polariton redshifts, which causes the loss of the polariton-laser coupling. Therefore, the emission falls back to the low intensity branch. 

Furthermore, by introducing noise on the laser power, we directly induce fluctuations on the polariton population, which, due to nonlinearity, bring fluctuations on polariton energy. Therefore, the presence of noise in the system disturbs the competition between feedback and loss mechanisms. This appears to be more sensitive close to the upper and lower threshold powers when the equilibrium between both mechanisms is vulnerable. As a consequence, the fluctuations induce a transition from one to other branch before attaining the threshold powers which causes the narrowing of the bistability width.
\subsection{A.2. Time-resolved experiments:}
\subsection{Residence time and Kramers time}
In order to investigate the changes of the bistability cycle we perform a time-resolved analysis of the polariton emission. We measure, at different input powers, the time behaviour of the polariton emission intensity for a wide range of noise strengths. In Figure 4 (a) to (d) we display the time streams of the polariton emission intensity recorded for different input laser powers, at an applied noise strength of 0.30 $\Delta$B. We observe that at 0.24 mW polaritons jump between the two stable states with similar probabilities. Note that, this excitation power corresponds to the power which the hysteresis collapses with 0.14 $\Delta$B noise strength. However, as the power is decreased to 0.20 mW, or increased to 0.28 mW, polaritons tend to become stable on the lower or on the upper state, respectively. We measure the time intervals spent in the lower (upper) branch before transition to the upper (lower) branch, and construct an histogram of the number of counts as function of time intervals. In Figure 5 we show the histogram for time intervals where polaritons stay in the lower branch for the case of an input power of 0.24 mW and applied noise of 0.48 $\Delta$B. This histogram gives an exponential decay with a characteristic time, called the residence time, of 17.5 $\mu$s (Inset of Fig. 5). Note that the residence time -the time period the system stays in each branch before it undergoes a transition- depends on the laser power and noise strength.

In Figure 6 (a) and (b), we plot, for different input laser powers, the residence time in the lower and upper branch as function of noise strength. It appears that for both branches, the residence times decrease as the noise strength increases. This is expected because, as we mentioned above, the input laser power fluctuations induce fluctuations in the polariton population, which produce instability in the hysteresis cycle. In the lower branch, the residence time decreases as the power is increased toward the upper threshold, and the reverse occurs in the upper branch. Note that, for 0.24 mW input power, the residence times for the lower and the upper branch coincide. This is the so called Kramers time \cite{Kramers1940}. It has to be pointed out that 0.24 mW is the power at which the hysteresis cycle collapses for 0.14 $\Delta$B noise (Figure 2) and also corresponds to the middle power of the hysteresis loops measured for lower amounts of noise (Figure 3 (b)). This result shows that, at this power, the polaritons have the same probability to stay at lower and upper bistable states. 
\begin{figure}[tb] 
\centering 
\includegraphics[width=1\columnwidth]{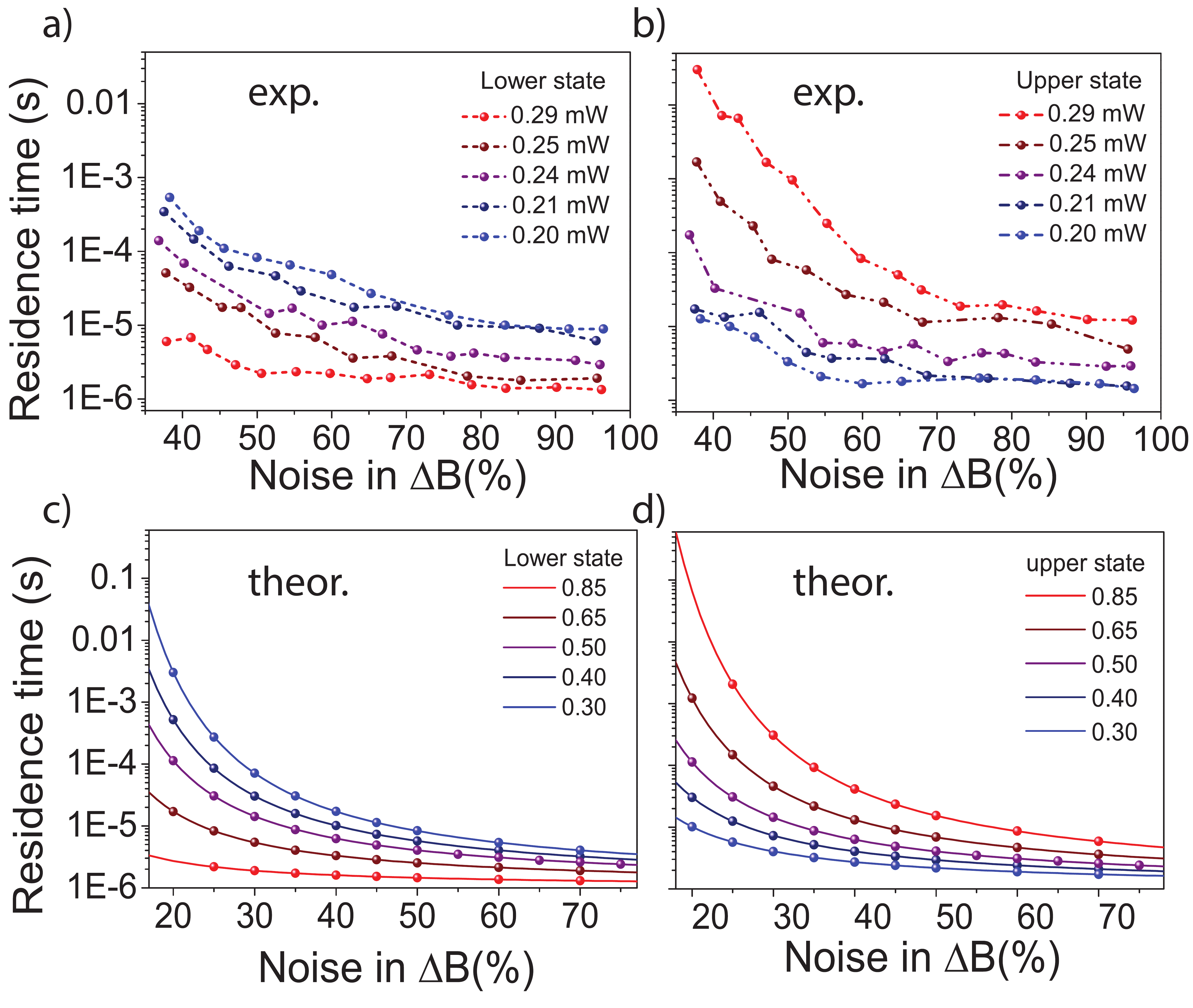} 
\caption[Polariton residence time for the lower and the upper states.]{\textbf{Polariton residence time for lower and upper states.} \textbf{a} Residence time of the lower state for five different laser powers : P=0.20, 0.21, 0.24, 0.25 and 0.29 mW. The noise intensity is increased from D=0.37 $\Delta$B to 0.96 $\Delta$B.  \textbf{b} Residence time of the upper state for the same amounts of laser power and noise. At P=0.24 mW residence times of the lower and the upper states are nearly the same. \textbf{c} Theoretical residence time of the lower state for five different laser powers :$P_{I}$=0.30, 0.40, 0.50, 0.65 and 0.85. The noise intensity is increased from D=0.2 $\Delta$B to 1 $\Delta$B.  \textbf{d} Theoretical residence time of the upper state for the same amounts of laser power and noise. Calculation from GPE (points), and theoretical model (line) are both presented.}
\end{figure}

We display in Figure 7 (a) the residence time in the upper and lower branch for a large variation of the noise strength measured at a power of 0.24 mW. This result shows that both residence times coincide showing that the probability to remain in the upper and lower branch is the same whatever is the strength of applied noise. It is worth mentioning that the minimum value measured for both residence times are determined by the noise correlation time (2$\mu$s). 

We would like to comment on the fact that the applied noise acts as an effective source of internal noise on the polariton system. This is shown to affect the overall character of the bistable system. It is worth mentioning the phenomenon of stochastic resonance \cite{Badzey2005}, which involves the counter-intuitive concept that, under certain conditions, adding noise to a modulated bistable system can actually increases its coherent response. Having established that the laser noise is a possible way of coupling external noise to the polariton system, it is indeed possible that, for the right amount of noise, one would see an increase in the coherent response of the amplified modulated signal, rather than the expected decrease. As a matter of fact, stochastic resonance has been observed in polariton bistable systems \cite{Abbaspour2014, Abbaspour2015}. It has been shown that, by applying in the input laser power a small amplitude modulation with frequency around 1 kHz, the stochastic resonance appears for a range of noise power (between 0.1 $\Delta$B to 0.3 $\Delta$B) in the polariton system \cite{Abbaspour2014}. Actually here we found that, for this amount of noise, the Kramers time is of the order of milliseconds, which is indeed around the modulation period as expected for the observation of stochastic resonance \cite{Wellens2004}. Notice that the hysteresis cycle appears as a discriminator for this amount of noise (Fig. 2 (a)). This effective noise value fixes the limit for the applied modulation frequency for the observation of stochastic resonance. 
\begin{figure}[t!] 
\centering 
\includegraphics[width=1\columnwidth]{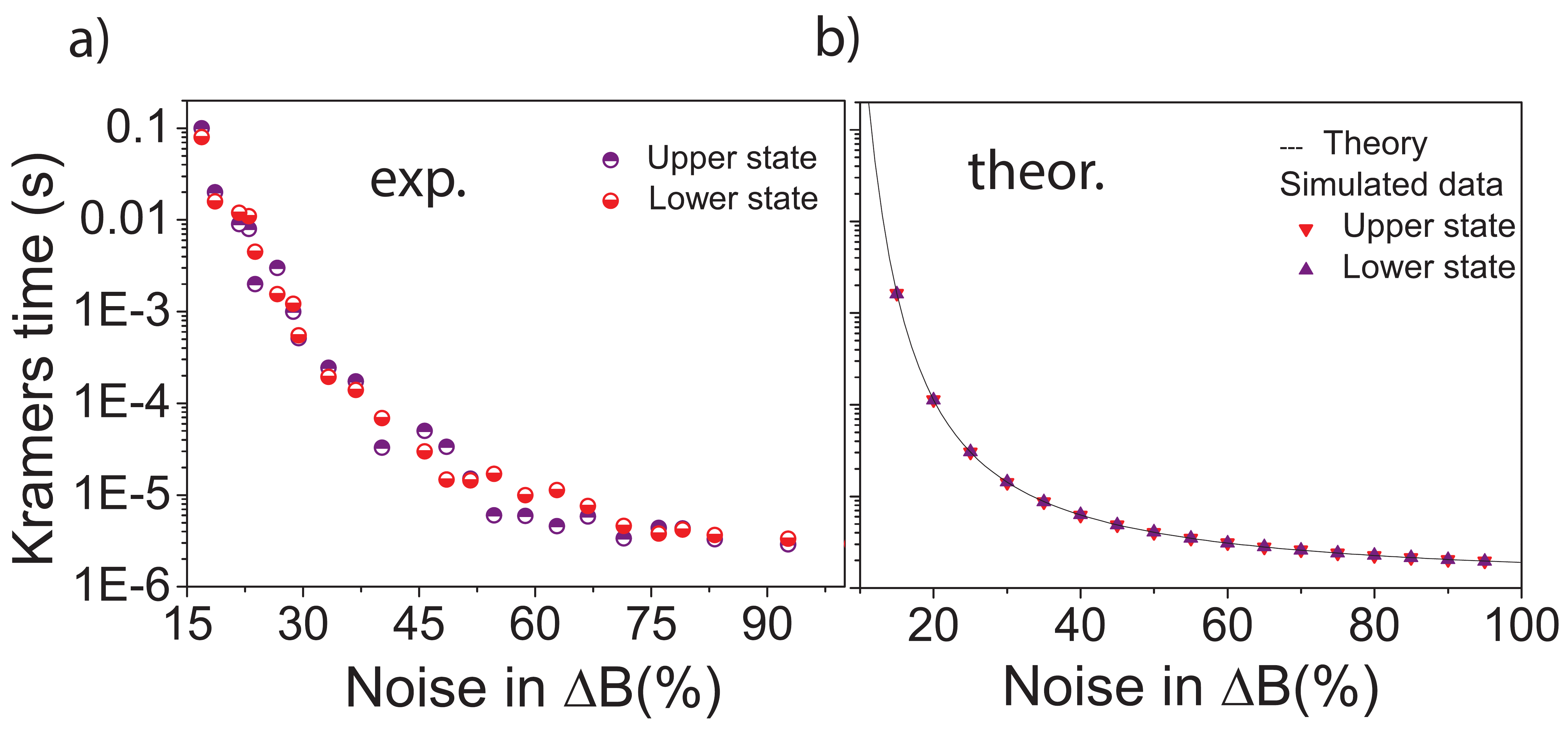} 
\caption[Polariton Kramers time.]{\textbf{Polariton Kramers time.} \textbf{a} Experimental polariton residence time for the lower (pink) and the upper (purple) states versus laser noise at the middle of polariton bistability (P=0.24 mW). At this position, the two residence times are nearly equal, which are then labelled as the Kramers time. The noise intensity is tuned from D=0.16 $\Delta$B to 0.93 $\Delta$B. When increasing the noise intensity, the Kramers time is observed to decrease. The minimum measured Kramers time is limited to the experimental bandwidth. \textbf{b} The theoretical residence times of the lower and the upper states for the laser power around the middle of bistability ($P_{I}$= 0.5), which is called as Kramers time. Calculation from GPE (points), and theoretical statistical model (line) are both presented.}
\end{figure}
\section{IV. THEORETICAL MODEL}
In order to model the polariton dynamics in a noisy bistability regime, we consider two different approaches. First, we perform numerical simulation using the Gross-Pitaevskii equation (GPE) with a stochastic perturbation. Second, starting with the polariton bistability without any external noise, we consider the statistical behaviour of the system.

For the numerical simulation we use the nonlinear GPE:
\begin{equation}
i \frac{d\Psi}{dt}=-\Delta\Psi-i\gamma_{p}\Psi+ \alpha_1|\Psi|^2\Psi+F
\end{equation}
where $\Psi$, $\gamma_{p}$ and $\alpha_{1}$ are the polariton field, the polariton linewidth and the interaction constant,  respectively. $\Delta$ represents the energy detuning between the laser and the polariton ground state energy. The driving field F is defined as:
\begin{equation}
F(t)=\sqrt{I+D(t)}
\end{equation}
where I is the laser intensity and D is the random perturbation term which acts on the intensity. The perturbation is considered as a Gaussian noise with a standard deviation $\sigma$. 
\subsection{Static regime}
In Figure 2 (c), (d) we display the simulated polariton bistability without noise (black dashed line) to extract the theoretical width $\Delta$B. We also display the bistabilities for noise with different standard deviation values normalised with respect to $\Delta$B. The parameters used for the GPE, to reproduce the experimental bistable behaviour, are $\gamma_{p}$=0.08 meV, $\alpha_{1}$=0.34 meV and $\Delta$=0.4 meV. We observe a qualitative agreement between the experimental (Fig. 2 (a, b)) and theoretical (Fig. 2 (c, d)) results. It is worth mentioning that the threshold power between the lower and the upper state depends on the characteristic time of the experiment compared to the residence time for a given noise strength, which is taken into account by the number of iterations in the simulation. 

In Figure 3 (c), (d) we display the simulated hysteresis width and the input power corresponding to the middle position of the respective hysteresis loops as a function of the applied noise strength, respectively. As clearly seen, the numerical simulations reproduce really well the three characteristic behaviours of the polariton dynamics in a noisy bistability regime: 1) the reduction of the bistability width for increasing noise power, 2) the linear dependence of this reduction and 3) the stability of the hysteresis middle point. It is worth mentioning that theoretical bistability width, without laser noise, is larger than corresponding experimental $\Delta$B. This discrepancy originates from mechanical or internal fluctuations which present during the experimental measurements.
\subsection{Dynamic regime}
Figure 4 (e) to (h) shows the simulated output time streams for different input powers and a fixed noise standard deviation D=0.20 $\Delta$B. The noise strength and input powers in GPE, are chosen to reproduce the experimental conditions. By using the same analysis procedure as for experimental data, we extract the residence time (Fig. 6 (c, d)) and Kramers time (Fig. 7 (b)).

To model the dynamics of the system between the lower and the upper states we also use a statistical approach according to the hysteresis loop without noise. Considering that the noise affects only the intensity and not the phase of the driving field, both thresholds on the hysteresis loop are fixed and perfectly defined. We name these intensity thresholds $I_{up}$ when the polaritons pass from the lower state to the upper state and $I_{down}$ for the opposite transition.

In this context, the noise perturbation D is modeled by a standard normal distribution with a characteristic correlation time ($\tau_{cor}$) given by the bandwidth of the experimental components. For a given mean power $I_{0}$ the noise intensity distribution becomes:
\begin{equation}
N(I)=\frac{1}{\sigma\sqrt{2\pi}}\exp{-\frac{1}{2}(\frac{I-I_{0}}{\sigma})^2}
\end{equation}
where $I_{0}$ and $\sigma$ are defined as a function of the $\Delta B$=$I_{up}-I_{down}$:
\begin{equation}
I_{0}=I_{down}+P_{I}\Delta B
\end{equation}
\begin{equation}
\sigma=P_{\sigma}\Delta B
\end{equation}
where $P_{I}$=0, 0.5 and 1 represent lower threshold, middle of optical bistability, and upper threshold, respectively. Finally, we can determine the conditional probability for the system to transit from the initial state by the cumulative distribution function (CDF). For example, when the polariton population is in the lower state, the probability to pass to the upper state is given by:
\begin{equation}
P(I>I_{up}\vert down)=\int_{I_{UP}}^{\infty} I(x) dx=\frac{1}{2}(1-\text{erf}(\frac{1-P_{I}}{P_{\sigma}\sqrt{2}}))
\end{equation}
where erf(x) is the error function. 

From a statistical point of view, the histogram built in Figure 5  represents the non-normalized probability to jump at time t knowing that the system was located on the lower state for t seconds. This probability function can be written as an analytical form:
\begin{equation}
W(n)=P_{s}^{(n-1)}P_{j}
\end{equation}
where $P_{j}$ is the probability to jump from initial state to the other stable state, and $P_{s}=1-P_{j}$ is the probability to stay in the initial state. n represents the number of iteration. Equation 7 can be written as:
\begin{equation}
W(n)= \frac{P_{j}}{P_{s}}\exp(n \text{ln}(P_{s}))
\end{equation}
Considering the noise correlation time ($\tau_{cor}$) as the minimum time step, this equation can be rewriten as:
\begin{equation}
W(t)= \frac{P_{j}}{P_{s}}\exp(\frac{t}{\tau_{cor}} \text{ln}(P_{s}))
\end{equation}
Similarly to the experimental procedure, we can extract the general residence time  ($\tau_{res}$):
\begin{equation}
\tau_{res}=-\frac{\tau_{cor}}{\text{ln}(P_{s})}
\end{equation}
and the particular Kramers time when
\begin{equation}
P(I>I_{up}\vert down)=P(I<I_{down}\vert up)=\frac{1}{2}(1+erf(\frac{0.5}{P_{\sigma}\sqrt{2}}))
\end{equation}
Considering the statistical theory, we add in Figures 6 (c, d) and 7 (b) the residence time and the Kramers time for different input laser powers as a function of the noise strength . The statistical description is in agreement both with the experiment and the GPE simulation. We can also extract certain experimental parameters such as the noise correlation time $\tau_{cor}$=1 $\mu$s which is close to the experimental value.
\section{V. CONCLUSION}
In conclusion, we have presented the noise dependence of the properties of a bistable polariton emission system. We have evidenced that the applied noise acts as an effective source of internal noise on the system. In the low noise regime, the hysteresis is well-defined and shows high-quality emission stability in each branch. We have shown that the upper (lower) threshold power decreases (increases) with noise strength, which induces polariton escape to the other branch. This results in reduction to the bistability width. We have demonstrated that a noise strength threshold  exist in which the bistability collapses and the system behaves as a discriminator. The results reveal that the threshold noise strength brings in lost of fidelity for devices. For larger amount of noise we have determined the residence times for different input powers through time resolved emission measurements. We have demonstrated that at collapsing power the probability of polaritons being in the upper and lower branch is the same whatever is the strength of applied noise. At this particular situation we have defined the Kramers time, which is measured as function of a large range of applied noise. Through numerical simulations using Gross-Pitaeviskii equation driven by stochastic excitation we have reproduced the experimental results. The experimental procedure and the theoretical background exposed in the present paper might be extended to other bistable structures. This paper present key elements to perform a study of noise induced phenomena in nonlinear systems, a prerequisite for device perfomances evaluation..
\section{ACKNOWLEDGMENTS}
The present work has been supported by the Swiss National Science Foundation under Project No. N135003, the Quantum Photonics National Center of Competence in Research under Project No. N115509, and by the European Research Council under Project POLARITONICS Contract No. N219120. The POLATOM Network is also acknowledged.

%

\end{document}